\documentclass[format=acmsmall, review=false, screen=true]{acmart}

\usepackage{booktabs} 

\usepackage{graphicx}           
\usepackage{amsmath}
\usepackage{multirow}
\usepackage{array}
\usepackage{geometry}
\usepackage{pdflscape}
\usepackage{rotating}
\usepackage{longtable, lscape, adjustbox}
\usepackage{tabularx}
\usepackage{subfigure}
\usepackage{appendix}

\usepackage[T1]{fontenc}

\acmJournal{CSUR}
\acmVolume{9}
\acmNumber{4}
\acmArticle{39}
\acmYear{2017}
\acmMonth{3}
\copyrightyear{2009}

\setcopyright{acmlicensed}

\acmDOI{0000001.0000001}

\received{February 2007}
\received[revised]{March 2009}
\received[accepted]{June 2009}

\begin{document}
\title[Understanding Application-level Caching]{Understanding Application-level Caching in Web Applications: a Comprehensive Introduction and Survey of State-of-the-art Approaches}  
\author{Jhonny Mertz}
\orcid{0000-0002-2522-4700}
\affiliation{%
  \institution{Universidade Federal do Rio Grande do Sul}
  \department{Instituto de Inform\'{a}tica}
  \city{Porto Alegre}
  \state{RS}
  \postcode{91501-970}
  \country{BR}
}
\author{Ingrid Nunes}
\affiliation{%
  \institution{Universidade Federal do Rio Grande do Sul and TU Dortmund}
  \department{Instituto de Inform\'{a}tica}
  \city{Porto Alegre}
  \state{RS}
  \postcode{91501-970}
  \country{BR}
}

\begin{abstract}
A new form of caching, namely application-level caching, has been recently employed in web applications to improve their performance and increase scalability. It consists of the insertion of caching logic into the application base code to temporarily store processed content in memory, and then decrease the response time of web requests by reusing this content. However, caching at this level demands knowledge of the domain and application specificities to achieve caching benefits, given that this information supports decisions such as what and when to cache content. Developers thus must manually manage the cache, possibly with the help of existing libraries and frameworks. Given the increasing popularity of application-level caching, we thus provide a survey of approaches proposed in this context. We provide a comprehensive introduction to web caching and application-level caching, and present state-of-the-art work on designing, implementing and managing application-level caching. Our focus is not only on static solutions but also approaches that adaptively adjust caching solutions to avoid the gradual performance decay that caching can suffer over time. This survey can be used as a start point for researchers and developers, who aim to improve application-level caching or need guidance in designing application-level caching solutions, possibly with humans out-of-the-loop.
\end{abstract}

%
%


%
%



\thanks{Author's addresses: J. Mertz {and} I. Nunes, Instituto de Inform\'{a}tica, Universidade Federal do Rio Grande do Sul, Porto Alegre RS, 91501-970, Brazil; emails: \{jmamertz, ingridnunes\}@inf.ufrgs.br}

\maketitle

\renewcommand{\shortauthors}{J. Mertz and I. Nunes}

\section{Introduction}
\label{sec:introduction}

Due to the increasing user demand for web-based applications, many optimization techniques have been proposed over the past years~\cite{Ravi2009} to allow web-based applications to deliver adequate performance. A popular technique to scale and improve the performance of this kind of application is \emph{web caching}, which consists of keeping data that are expected to be frequently requested easier to be retrieved, e.g.\ without the need for recurrent calculations and processing.

Many efficient web caching solutions have been proposed and employed in the web system architecture. However, traditional caching solutions are deployed outside the application boundaries as a transparent layer, and thus caching decisions are not made taking into account particularities of specific web applications. As a result, traditional types of caching are usually less efficient when the web application processes complex logic and personalized content. Therefore, application-level caching can complement other caching solutions and potentially improve performance and scalability of web-based applications, as well as reduce the workload and communication delays of these systems.

However, application-level caching solutions are usually developed in an \emph{ad-hoc} way, based on conventional wisdom of web usage patterns~\cite{Mertz2016}. It means that the caching design and implementation, which may initially satisfy desired non-functional requirements, may become obsolete due to application changing dynamics. Thus, to ensure that the cache continues contributing to the application performance, a frequent adjustment of caching decisions is needed, which implies additional time spent by developers on maintenance~\cite{Radhakrishnan2004}.

Given the issues involved with application-level caching, substantial advances have been made towards supporting developers while developing such caching solutions. In this paper, we provide a comprehensive overview and comparison of existing approaches proposed in this context, so that it is possible to understand what can be put into practice and remaining open issues. We first provide an introduction to web and application-level caching and then focus on surveying static and adaptive approaches in the literature. Thus, the scope of this survey has two key dimensions: static and adaptive application-level caching approaches. The former refers to non-adaptive solutions to help developers design, implement and maintain an application-level caching solution, typically focusing on raising the abstraction level of caching, with the provision of caching implementation support or automating some of the required tasks, thus providing caching management support. The latter is focused on cache management approaches that can adaptively deal with caching issues, usually by monitoring and automatically changing behavior to achieve desired objectives.

Web caching is an in-depth studied optimization technique, and previous reviews on caching locations and deployment schemes~\cite{Ravi2009,MZhang2015,Abdullahi2015,Hattab2015,Tamboli2015}, coordination tasks~\cite{Podlipnig2003,Balamash2004,Ali2011}, measurement~\cite{Domenech2006}, and adaptation~\cite{Venketesh2009,Ali2011,SarinaSulaiman2011,Ali2012,Ali2012a} have already been published and discussed. However, these reviews mainly address the context of web pages (at the proxy level), which are not always best suited to the application-level or other caching locations. Although our paper may contain some overlapping topics, our survey differs itself from previous surveys because it focuses specifically on caching issues applied to application-level caching design, implementation, maintenance, and adaptation, which are associated with distinguished challenges.

The remainder of this paper is organized as follows. Section~\ref{sec:applicationlevel} gives a comprehensive overview of challenges and issues of application-level caching. Section~\ref{sec:approaches} presents static and adaptive application-level caching approaches, focusing on the issues addressed, benefits and techniques employed. Finally, conclusion, open challenges and future directions in the field are presented in Section~\ref{sec:conclusion}.

\section{Introduction to Application-level Caching}
\label{sec:applicationlevel}

As opposed to caching alternatives that are placed outside of the application boundaries, application-level caching allows storing content at a granularity that is possibly best suited to the application. For instance, modern web applications nowadays provide customized content and, in this case caching the final web pages is usually useless. We consider a web-based application any application that contains data and business logic maintained by developers or contains a presentation logic to provide content or features to users through the web. Therefore, application-level caching can be used to separate generic from specific content at a fine-grained level.

Application-level caching is mainly characterized by caching techniques employed along with the application code, i.e.\ business, presentation and data logic. Therefore, it is not tied up to a specific caching location (server-side, proxy or client-side), because it can be conceived at the server-side, to speed up a Java-based application that produces HTML pages sent to users, as well as at the client-side, as a JavaScript-based application that executes part of its logic directly on the client's browser. In both situations, developers can reason about caching and implement a caching logic to satisfy their needs.

Thus, application-level caching has become a popular technique to reduce the workload on content providers, in addition to other caching layers surrounding the application. Such popularity is confirmed by \citeN{Mertz2016} that analyzed ten web applications and showed that application-level caching represents a significant portion of lines of code of investigated applications as well as a significant number of issues, considering that caching is essentially a non-functional requirement. In addition, \citeN{Selakovic2016} analyzed 16 JavaScript open source projects and reported that 13\% of all performance issues are caused by \emph{repeated execution of the same operations}, which could be solved by means of application-level caching resulting in an improved performance and decreased user perceived latency.

We next provide an introduction to application-level caching. For a background on caching and web caching, we refer the reader to Appendix~\ref{sec:appendix:background}.

\subsection{Overview}

\begin{figure}
  \centering
  \subfigure[Process Steps.]{\includegraphics[width=0.4\linewidth]{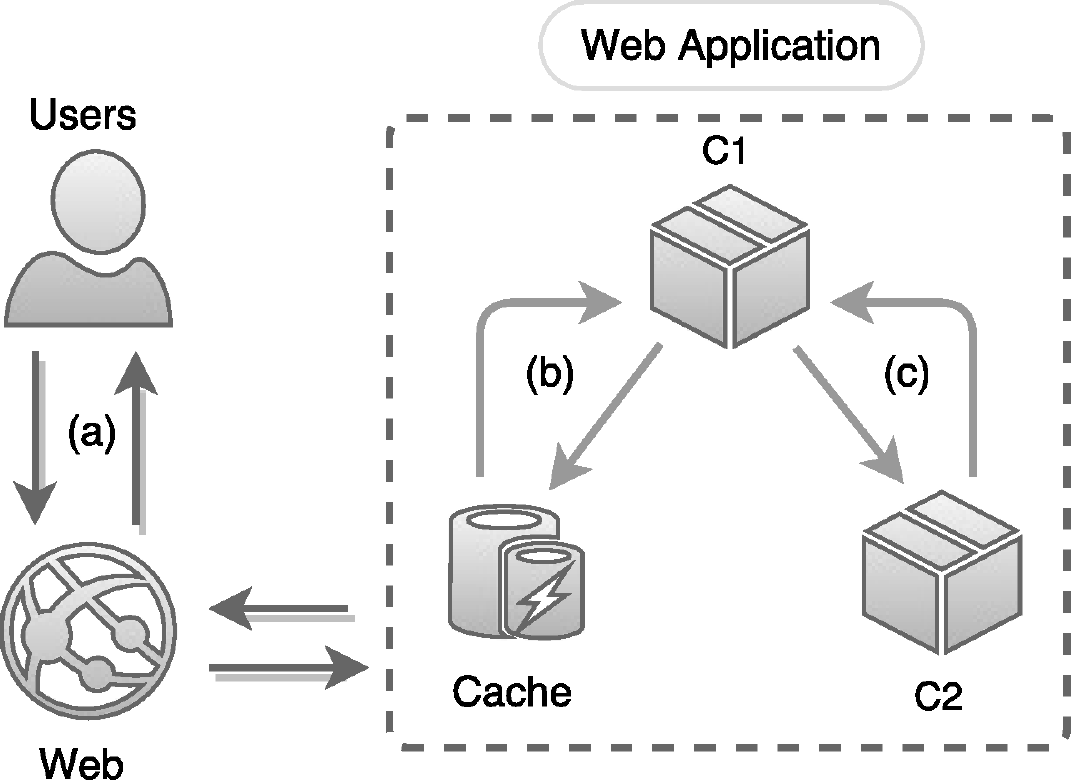}
  \label{fig:applicationcaching}}
  \subfigure[Code Example.]{\includegraphics[width=0.4\linewidth]{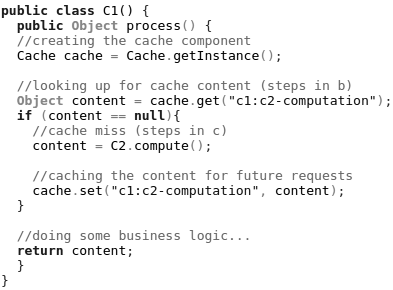}
  \label{fig:codeExample}}
  \caption{Application-level Caching Overview.}
  \label{fig:AppCacheOverview}
\end{figure}

To ease the understanding of what application-level caching is, we provide an introductory example in Figure~\ref{fig:AppCacheOverview}, in which the process of using and implementing application-level caching is detailed. First, a web application receives a request (\emph{step a}), which is eventually processed by a component C1. However, C1 depends on C2, and calling and executing C2 may imply an overhead regarding computation or bandwidth. Therefore, C1 manages to cache C2 results and, for every request, C1 verifies whether C2 should be called or there are previously computed results already in the cache (\emph{step b}). If the content is cached, a \emph{hit} occurs, and C1 can avoid calling C2. However, when a \emph{miss} occurs, computation by C2 is required (\emph{step c}). Finally, the results of C2's computation can be cached to process future requests faster. These steps are those commonly performed by any implementation of caching. The key difference is that, in application-level caching, the responsibility of managing the caching logic is entirely left to application developers, possibly with the support of frameworks that provide an implementation of the cache system, such as EhCache\footnote{\url{http://www.ehcache.org/}} and Memcached\footnote{\url{https://memcached.org/}}.

Given that application-level caching is essentially a caching solution, it shares commonalities such as coordination issues and metrics with other caching and web caching solutions. For more information regarding general caching and web caching issues, we refer the reader to Appendix~\ref{sec:appendix:background}, where we provide an extensive background on caching and web caching, pointing out to other surveys and papers that provide more detailed research on each topic.

\subsection{Challenges and Issues}
\label{subsec:applevelchallenges}

Building an application-level caching solution involves four key challenging issues: choosing \emph{what} content to cache, defining \emph{when} the selected content will be inserted into and removed from the cache component, determining \emph{where} to store cache content efficiently, and deciding \emph{how} to properly implement caching logic~\cite{Mertz2016}. The latter should consider that both the cache and underlying source of data should be handled and linked with each other by the application, as shown in Figure~\ref{fig:applicationcaching}. Therefore, developers must manually insert and retrieve content, translate between raw data and cache objects, assign keys, and keep consistency between the cache and the source of the cached content. Such logic is placed within the application and is usually tangled with the business logic---see Figure~\ref{fig:codeExample}, in which caching decisions are made explicit (i.e.\ which objects to get, put or remove from the cache), as opposed to an implicit cache, in which the cache is implemented as a transparent layer.

Caching implementation and maintenance are thus a challenge because caching becomes a \emph{cross-cutting concern} mixed with business logic and spread all over the application base code, resulting in increased complexity and maintenance time~\cite{Ports2010}. Moreover, web applications are usually not conceived to use caching since their beginning. As the application evolves, complexity or usage analysis is performed and may lead to a performance improvement demand~\cite{Radhakrishnan2004}. Developers must thus \emph{refactor} the application business code to insert caching logic into the proper locations. Although it is often impossible to predict that an application will require application-level caching in the future, posterior cache implementation can lead to significant rework due to refactoring, and an implementation that would have better quality if thought before the application reaches advanced stages of development.

Despite implementation issues, deciding \emph{what} and \emph{when} to cache specific content demand a significant effort and reasoning from developers. Such decisions involve the \emph{choice for cacheable content}, in which an admission policy should be adopted, and the definition of rules to \emph{maintain consistency} between cache and source of data, in order to avoid stale content. If these decisions are not taken properly, it can result in an increased cache memory consumption and, at the same time, the cached content does not lead to hits, which tends to decrease the application performance. Therefore, it is crucial to understand what are the typical usage scenarios, how often the data selected to be cached is going to be requested, how much memory this data consumes, and how often it is going to be updated. Finally, developers should concern \emph{where} the cached data will be stored. Such issue involves the management of the cache system, which consists of several \emph{non-trivial decisions}, such as establishing a replacement policy and defining the size of the cache~\cite{Mertz2016}.

These issues require a significant and manual effort from application developers, given that the design and maintenance of this caching demand extensive knowledge of the application to be properly done. Given these shortcomings, application-level caching development is not trivial and, therefore, is a time-consuming and error-prone task, as well as a common source of bugs~\cite{Ports2010,Gupta2011,Mertz2016}.

\subsection{Static vs.\ Adaptive Application-level Caching}

A fundamental problem of application-level caching is that all issues mentioned above usually demand extensive knowledge of the application to be properly solved. Consequently, developers manually design and implement solutions for all those mentioned tasks. To provide solutions to deal with caching that require less intervention, thus easier and faster to be adopted, static approaches have been proposed to help developers while designing, implementing and maintaining an application-level caching solution. \emph{Static approaches} usually focus on decoupling this non-functional requirement from the base code and providing ready-to-use caching components with basic functionalities and default configurations, automating some of the required tasks such as a storage mechanism and standard replacement policies. However, even when leveraging static solutions to ease the development of a caching solution, the issues and challenges related concerning design and maintenance remain unaddressed, because default configurations do not cover all the design issues, and those provided may not even perform well in all contexts. 

Therefore, developers must still specify and tune cache configurations and strategies, taking into account application specificities. A common approach to develop a cache solution is to define configurations and strategies according to well-accepted characteristics of access and common-sense assumptions. Then, cache statistics, such as hit and miss ratios, can be observed and used to improve the initial configuration. This tuning process of caching configurations is repeated until a steady performance improvement is achieved. As a result, to achieve caching benefits so that the application performance is improved, it is necessary to tune cache decisions constantly. Despite the effort to do so, eventually, an unpredicted or unobserved usage scenario may emerge. As the cache is not tuned for such situations, it would likely perform sub-optimally~\cite{Radhakrishnan2004}.

This shortcoming motivates the need for \emph{adaptive caching solutions}, which could overcome these problems by adaptively adjusting caching decisions at run-time to maintain a required performance level. Moreover, an adaptive caching solution minimizes the challenges faced by developers, requiring less effort and providing a better experience with caching for them. Such adaptive caching solutions aim at optimizing the usage of the infrastructure, in particular for the caching system.

Given that we introduced application-level caching and provided the background needed to understand the approaches that are discussed in this paper, we now proceed to the presentation of different ways to provide a caching solution for developers as well as static and adaptive approaches that were proposed in the context of application-level caching.

\section{Application-level Caching Approaches}
\label{sec:approaches}

\begin{figure}
  \centering
  \includegraphics[width=\linewidth]{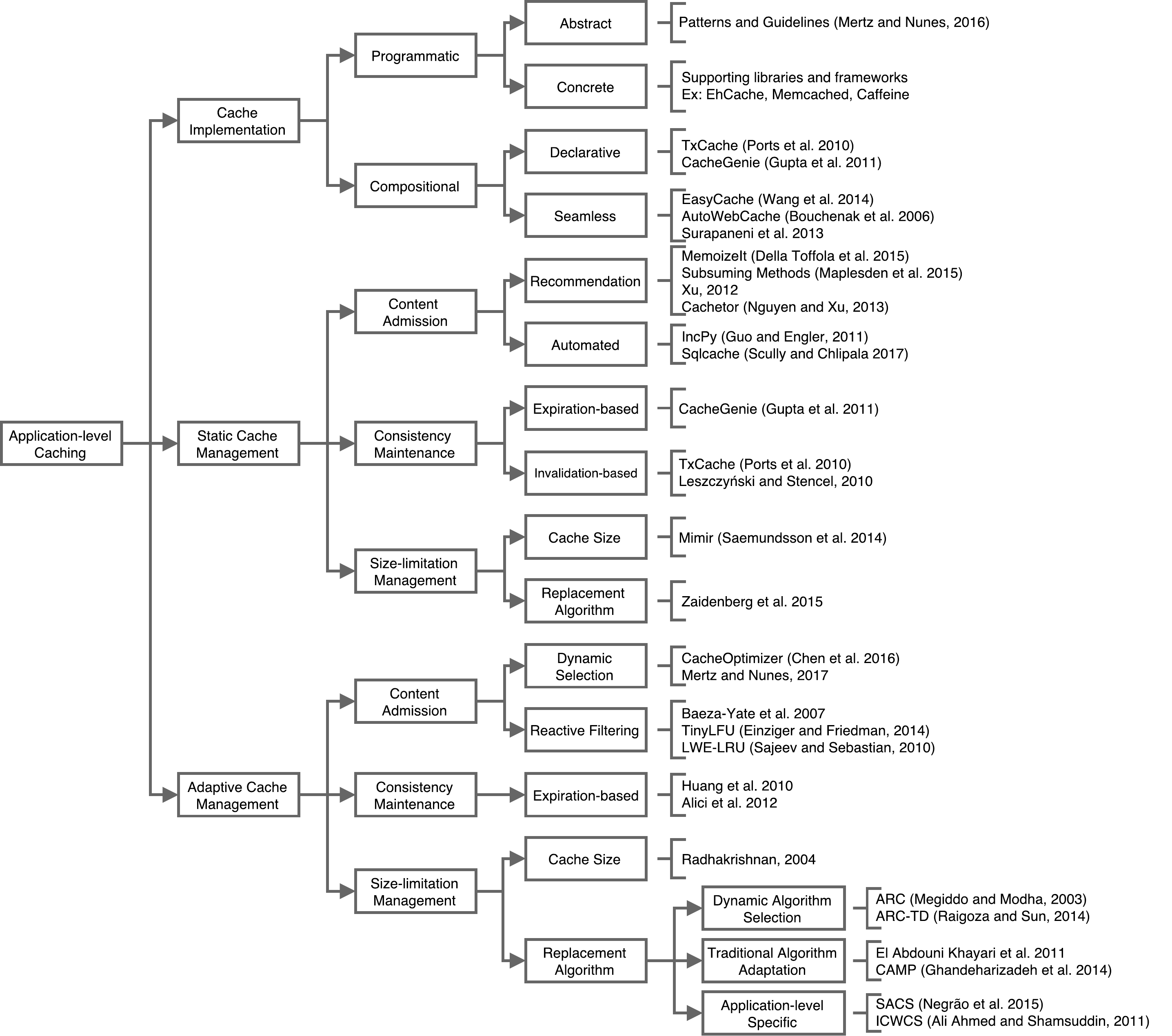}
  \caption{Classification of Application-level Caching Approaches and Representative Examples.}
  \label{fig:approaches-taxonomy}
\end{figure}

To introduce existing approaches in the context of application-level caching, we outline in Figure~\ref{fig:approaches-taxonomy} a taxonomy of application-level caching solutions, also indicating all the works discussed in our survey. Issues and challenges that appear in this taxonomy were introduced in the previous sections.

It is important to mention that our initial intention was to perform a systematic review of application-level caching approaches. However, while specifying appropriate search strings, many alternative searches led to either a small set of papers, with many important papers not being retrieved, or a large set that included a significant amount of unrelated works. This large set was unfeasible to be processed in a timely fashion. This is due to the investigation of caching in a wide variety of contexts. Given the different types of caching that can be employed through the web infrastructure and the lack of specific nomenclature that identifies each web caching technique, it is difficult to match solely papers that deal with application-level issues. Furthermore, terms such as \emph{adaptive}, \emph{web application}, and \emph{cache}, associated with our study, are widely used in many other contexts. As result, proceeding with a systematic approach would result in a poor review of application-level caching. Therefore, we conducted our survey with studies we collected using a less rigorous selection process. We collected many relevant papers using alternative query strings, and further searched for papers published by key authors or cited in relevant papers, using a snowball approach, until we reached a fixed point. This gives us confidence that relevant papers are indeed included in our survey.

As presented in Figure~\ref{fig:approaches-taxonomy}, approaches are classified into three categories, depending on how the approach helps developers with caching issues. We next discuss these approaches, following such categorization. First, we present work focused on supporting the implementation of a caching solution. Then, static approaches regarding coordination issues of application-level caching are presented and discussed. Finally, we introduce caching approaches that can adapt their behavior, typically by means of a feedback loop.

\subsection{Cache Implementation}

In order to reduce the effort demanded by developers while implementing application-level caching, implementation-centered approaches have been extensively developed. Such approaches focus on the provision of solutions that raise the level of abstraction and reduce a significant amount of caching logic to be added to the base code. Manually controlling and maintaining the caching logic may be error-prone and tedious, because it is often repetitive and not related to the business logic, as presented in Section~\ref{sec:applicationlevel}. Therefore, a research challenge in application-level caching is how to \emph{ease the implementation} of a caching logic for developers.

Usually, the idea underlying solutions to implementation issues is that the application effort can be reduced by providing a system or library that handles some caching operations, freeing developers to write the most relevant code (i.e.\ business logic). As shown in Figure~\ref{fig:approaches-taxonomy}, approaches related to caching implementation can be categorized into two types, concerning the adopted programming model: \emph{programmatic} and \emph{compositional}. The former model requires code changes to take advantage of the caching approach and usually results in an application-specific solution. For example, EhCache provides a full-featured caching component, which is responsible for dealing with memory and disk store. However, it requires coding with the EhCache provided classes or annotations to perform the caching operations. The latter has a lower impact on the application, as it does not require to introduce code interleaved with its base code.

\subsubsection{Programmatic}

The simplest programmatic approach for application-level caching consists of the use of \emph{abstract} solutions such as design and implementation patterns. Recent work done by \citeN{Mertz2016} provided practical guidance to developers as patterns and guidelines to be followed while designing, implementing and maintaining application-level caching, derived from a set of existing web applications that use application-level caching. Although simple, patterns are abstract solutions and require a significant amount of changes in the base code to be adopted, which may not provide an adequate degree of transparency and flexibility to developers~\cite{Pohl2005}. To provide \emph{concrete} support, libraries and frameworks have been developed, providing useful and ready-to-use cache-related features. Redis\footnote{\url{https://redis.io/}}, Memcached, Spring Caching\footnote{\url{https://docs.spring.io/spring/docs/current/spring-framework-reference/html/cache.html}}, EhCache and Caffeine\footnote{\url{https://github.com/ben-manes/caffeine}} are popular examples of libraries and frameworks. For a deeper analysis of typical cache systems and libraries, as well as their main techniques regarding data management, we refer the reader to elsewhere~\cite{MZhang2015}.

Typically, concrete solutions provide an in-memory storage component along with methods that allow to put and get arbitrary content, identified by a key, i.e.\ a hash table. The key benefit of such implementations is that they are: (a) flexible, because different types of content can be stored, ranging from database query results to entirely generated web pages; (b) simple; and (c) scalable. Besides being storage units, these caching solutions also provide support to basic caching issues, such as memory allocation, eviction policies, and size limitations. Despite these advantages, adopting such solutions still require a programming effort to manage the cache and couples caching concerns to the base code, which adds complexity and reduces the possibility of reuse within it~\cite{Ports2010,Wang2014}.

Distributed key-value stores (or caches) have become popular for supporting large-scale web applications because they can easily scale up---i.e.\ they can be distributed to multiple servers, which means it can linearly scale to handle greater transaction loads---and cache a significant portion of user requests, improving hit ratios and response time. Such distributed caches are an essential component in the infrastructure required by large enterprises such as Netflix\footnote{\url{http://techblog.netflix.com/2016/03/caching-for-global-netflix.html}}, Facebook~\cite{Nishtala2013} and Twitter\footnote{\url{https://github.com/twitter/twemproxy}}. Besides design, implementation and maintenance issues of application-level caching, there is research work that proposes and improves cache components in the context of distributed infrastructures~\cite{Zhu2012,Fan2013,Xu2014,Li2015a}. These approaches are not further discussed in this survey, given that our focus is on issues specific to application-level caching, and not broad caching issues, such as storage, concurrency management, and other topics of distributed systems. Such topics have been widely researched and addressed by other communities~\cite{HZhang2015,Abdullahi2015}.

Applications can also be developed with business logic at the client-side, typically with the support of JavaScript frameworks. Popular frameworks provide features to develop complex and sophisticated applications. However, caching configurations are usually specified by the content provider (web server that received the request) in the header of the response, and caching is automatically done by browsers, which are limited to caching cache entire web pages or request responses. To increase the flexibility regarding the customization of caching configurations at the client-side, \citeN{Huang2010} proposed a caching framework in which developers can define the appropriate strategy among pre-defined possibilities. Similar to the caching provided by browsers, the proposed framework does not require developers to manage the cache content manually but allows customizations, such as setting cache granularity and eviction policies. \citeN{Huang2010} also proposed an approach to adaptively deal with consistency management, which is better explored in Section~\ref{sec:adaptiveapproaches}.

\subsubsection{Compositional}

\emph{Compositional} approaches can partially relieve the developers' burden by automating tasks through \emph{declarative} or \emph{seamless} approaches. The former is based on the provision of knowledge associated with the semantics of application code and data. In this case, annotations referred to as \emph{assertions} or \emph{contracts}, are used to describe application properties, which can be processed at runtime to execute specific tasks without coupling the caching logic with the application base code. Such annotations have become widely adopted in dynamic programming languages~\cite{Stulova2015}. \emph{CacheGenie}~\cite{Gupta2011} is a system that provides a higher level of abstraction for caching frequently observed query patterns in web applications. These abstractions take the form of declarative query objects and, once developers specify them, insertions, deletions, and invalidations are done automatically. CacheGenie is based on triggers inside the database to automatically invalidate the cache, or keep it synchronized with the data source, as expressed by the developer in the application code. Similarly, \citeN{Ports2010} proposed \emph{TxCache} which offers for developers a simple way to indicate database query functions as cacheable, and then TxCache automatically manages and caches those marked function results. CacheGenie and TxCache also provide automatic consistency management, which is better explored in Section~\ref{sec:staticapproaches}.

Seamless solutions are those that are coupled to the application in a transparent way, being added to the application, for example, as a surrounding layer, similarly to, e.g., a database and web proxy, without the need for refactoring application code. Such solutions are especially useful in scenarios in which the application was not conceived to use caching since the beginning, and requests for performance and scalability improvements emerged after many releases. In this context, middleware-based approaches that integrate two or more layers of a web application have been proposed as easy-to-use and transparent solutions to deal with caching~\cite{Huang2010,Wang2014}. Furthermore, aspect-oriented programming (AOP)~\cite{Kiczales1997} has been explored by increasing modularity with an improved separation of cross-cutting concerns. As caching is fundamentally a cross-cutting concern, \citeN{Bouchenak2006} and \citeN{Surapaneni2013} demonstrated the use of AOP to develop caching solutions. Such approaches come as an alternative to refactoring the application with the introduction of programmatic approaches. Although faster results may be obtained with transparent solutions, they can be hard to support and tune, because system administrators, or even developers, might need to be specially trained or experienced in particular caching solutions and scenarios to configure them properly.

A non-exhaustive survey on database and mid-tier transparent approaches are presented by \citeN{Ravi2009}, while \citeN{Ali2011} surveyed some studies of proxy-level caching. We complement such surveys with seamless approaches that consider application-level specificities, which are better explored in Sections~\ref{sec:staticapproaches} and \ref{sec:adaptiveapproaches}, given they also provide static or adaptive solutions to design and maintenance of application-level caching.

Although programmatic and compositional approaches focused solely on implementation issues can raise the level of abstraction and, consequently, reduce a significant amount of caching logic to be added to the base code, they still require design reasoning, such as deciding whether to cache content and ensuring consistency. Therefore, a more fine-grained solution would require reduced effort and input from developers, which are the focus of the approaches described next.

\subsection{Static Cache Management}
\label{sec:staticapproaches}

Static application-level approaches are those that address design and maintenance issues of caching focusing on automating required tasks or giving suggestions towards easing the developer reasoning. According to the taxonomy presented in Figure~\ref{fig:approaches-taxonomy}, static cache management solutions are classified into three categories, depending on which caching issue they address. The remainder of this section groups static approaches according to these categories.

\subsubsection{Content Admission}

There are two main ways of helping developers select cacheable content: providing caching \emph{recommendations}, or providing an \emph{automated} admission by automatically identifying and caching content. Table~\ref{tab:static-admission-approaches} summarizes the caching approaches that deal with admission issues.

\begin{table}
\scriptsize
\centering
\renewcommand\tabcolsep{1mm}
\caption{Static Caching Approaches for Selection of Cacheable Content.}%
\begin{tabular}{|c|p{1.5cm}|p{1.7cm}|p{1.6cm}|p{2.2cm}|p{2.5cm}|p{2.6cm}|}
\hline
 & \textbf{Approach} & \textbf{Based on} & \textbf{Content} & \textbf{Data Input} & \textbf{Analysis} & \textbf{Output} \\ \hline

\multirow{4}{*}[-11ex]{\rotatebox[origin=c]{90}{Recommendation}}
  & MemoizeIt \cite{DellaToffola2015} & Iterative profiling & Method calls & Time, frequency, and input-output profiling & Hit ratio, invalidation and size thresholds, and estimations & Report with ranked list of potential opportunities to the user for manual inspection \\ \cline{2-7}

  & Subsuming Methods \cite{Maplesden2015} & Application profiling & Method calls & Calling context tree & Customized metric based on method distance in the context tree & Subset of the methods in the application that are interesting from a performance perspective \\ \cline{2-7}

  & \citeN{Xu2012} & Application profiling & Data{\par}Structures & Heap data structures for each allocation site during the execution of the application & Customized metric to approximate reusability based on three different levels (instance, shape, and data) & Report with a list of top potentially reusable allocation sites to the user for manual inspection \\ \cline{2-7}

  & Cachetor \cite{Nguyen2013} & Abstract dependency graph analysis & Bytecode instructions, data structures, and method calls & Instructions, data structures and call sites profiling & Cacheability measurements for each type of content based on frequency & Ranked list of potential opportunities \\ \hline

\multirow{5}{*}[-11ex]{\rotatebox[origin=c]{90}{Automated Caching}}
  & IncPy \cite{Guo2011} & Application profiling & Method calls & File accesses, value accesses, and method calls & Safeness (consistency) and worthwhileness (expensiveness) heuristics & Automatically caches and invalidates data files \\ \cline{2-7}
  
  & Sqlcache \cite{Scully2017} & Application profiling & Database queries generated by the application & SQL statements & SQL analysis and program instrumentation & Automatically caches and invalidates database query results \\ \cline{2-7}
  
  & EasyCache \cite{Wang2014} & Database queries monitoring & Simple database queries & SQL statements & Query text analysis & Automatically caches and invalidates database query results \\ \cline{2-7}
  
  & AutoWebCache \cite{Bouchenak2006} & Web requests and database queries monitoring & Final web pages & Servlet method calls and definition of database queries & Content dependency analysis & Automatically caches and invalidates web pages \\ \cline{2-7}
  
  & \citeN{Surapaneni2013} & Database queries monitoring & Database sub-queries & Frequency, cost of updates and selectivity of database queries & Utility function-based & Automatically caches and invalidates sub-queries \\ \hline
  
\end{tabular}
\label{tab:static-admission-approaches}
\end{table}

The first way of helping developers while admitting content in the cache is by recommending improvement opportunities. Approaches in this context are usually based on the analysis of application profiling information, which can capture application-specific details through monitoring its execution when facing different situations. Traditional profiling approaches typically record measurements of method calls. Usually, cost measurements are captured with calling context information, which conveys the hierarchy of active methods calls of a request.

\citeN{DellaToffola2015} addressed this problem by identifying and suggesting method-caching opportunities. Their approach, called \emph{MemoizeIt}, is based on comparing inputs and outputs of method calls and dynamically identifying redundant operations, according to specified thresholds. To prevent the expected overhead of the approach, which implies a comparison of all method invocations, MemoizeIt analyzes the collected executions level by level through iterations. First, it compares input and output objects without tracking dependencies, and then iteratively tracks deeper dependency levels of objects that remain under the specified thresholds. By doing this, MemoizeIt can explore the application traces faster to define a set of methods that would provide benefits if cached. Also by analyzing method calls in an application profile, the work of \citeN{Maplesden2015} can identify the entry point to repeated patterns of method calls, which are called \emph{subsuming methods} and are identified by analyzing the smallest parent distance among all the common parents of a method.

\citeN{Xu2012} focused on a common problem in object-oriented applications, in which an allocation site creates data structures with the same content repeatedly, but in different moments during an execution. He follows the same principle of helping developers with a report, which lists the top allocation sites that create such data structures. Then developers can manually inspect the code and implement the appropriate solution to improve performance. \emph{Cachetor}, proposed by \citeN{Nguyen2013}, addresses repeated computations by suggesting spots of invariant data values that could be cached for later use. They proposed a runtime profiling tool, which uses a combination of dynamic dependency and value profiling to identify and report operations that keep generating identical data values. However, Cachetor makes strong assumptions about the programming language, such as the presence of a specialized type system, which is not valid in dynamically-typed languages. \citeN{Infante2014} addressed this type system restriction by proposing a multi-stage profiling technique that uses dynamically collected data to reduce the profiling overhead.

Although these approaches can reduce complexity and time required by caching design, developers should still review the recommendations, decide whether to cache or not the suggested opportunities and integrate the appropriate caching logic into the application. Thus, a second way of helping developers while dealing with content admission is to automatically identify and cache the cacheable content, as opposed to just report potentially cacheable methods. Such approaches require not only ways to analyze the application behavior, but also mechanisms to manage cache and application at runtime.

\citeN{Guo2011} achieved this by implementing a customized Python interpreter, namely \emph{IncPy}, which includes an approach to identify and cache repetitive creation or processing of data files stored on disk. Such disk accesses may lead to long-running method calls resulting in bottlenecks. Thus, at runtime, it monitors and analyzes disk operations based on heuristics regarding safeness and worthiness, which are used as reference to cache and evict content. Similarly, \citeN{Scully2017} proposed \emph{Sqlcache}, which consists of a compiler optimization that is able to analyze the application code and generate the appropriate caching logic for database queries. However, Sqlcache is specifically proposed and implemented to web applications in the Ur/Web\footnote{\url{http://www.impredicative.com/ur/}} programming language.

\emph{EasyCache}~\cite{Wang2014} combines properties of database caching with application-level specificities to provide a mid-tier mechanism, which caches database queries and translates it into application-level objects in a seamless way. Such approach is built on top of a popular application programming interface to access databases, and thus does not require any modification of the application code. As it is delivered as a caching layer over the integration between application and database, it is able to inspect all the text from database queries that comes from the application in order to automatically find cacheable queries and maintain consistency. However, in order to keep the approach feasible regarding memory and network bandwidth, complex queries (e.g.\ nested and statistics queries) are unsupported, as well as large objects. Moreover, because it is implemented at a lower level, as a Java Database Connectivity (JDBC) driver, the approach does not work with object-relational mapping frameworks (despite the authors argued that the integration is feasible), which are commonly used while implementing web applications.

Also based on JDBC, \citeN{Bouchenak2006} used AOP to implement a middleware support for caching, named \emph{AutoWebCache}. Such middleware is based on intercepting web application servlets as well as database queries through JDBC. AutoWebCache then automatically integrates front-end and back-end by analyzing the queries to identify cacheable web pages. Furthermore, such analysis gives the content dependencies, which are used to trigger cache invalidations to maintain consistency with the back-end database. Despite built to support only databases as source of information, this approach is generic enough to be implemented with other types of sources. Furthermore, AOP is used by \citeN{Surapaneni2013}, which described an approach to cache the results of join sub-queries, with the goal of caching partial query results, rather than whole queries. Such approach identifies cacheable sub-queries based on a cache factor, which takes into account monitored information such as frequency, cost of updates and an estimation of selectivity (calculated through histograms built from observed collections of data at runtime) of sub-queries.

\subsubsection{Consistency Maintenance}

We can classify static application-level caching consistency approaches as relying on two main techniques: \emph{expiration} and \emph{invalidation}. The former provides higher availability by allowing stale data to be returned from the cache. The latter ensures that no stale data is returned by identifying when the underlying data source is modified and evicting related cached items. For more detailed information regarding web caching consistency concepts, please refer to Appendix~\ref{sec:appendix:background}. Table~\ref{tab:static-consistency-approaches} summarizes the caching approaches that address these problems of dealing with consistency management.

\begin{table}
\scriptsize
\centering
\renewcommand\tabcolsep{1mm}
\caption{Static Caching Approaches for Consistency Management.}%
\begin{tabular}{|p{2cm}|p{2.1cm}|p{1.8cm}|p{2.5cm}|p{2.1cm}|p{2cm}|}
\hline
\textbf{Approach} & \textbf{Based on} & \textbf{Content} & \textbf{Data input} & \textbf{Analysis} & \textbf{Consistency}{\par}\textbf{Technique} \\ \hline

TxCache \cite{Ports2010} & Database queries monitoring & Results of database queries generated by application & Queries designated as cacheable by developers & Transactional context & Invalidation-based \\ \hline

CacheGenie \cite{Gupta2011} & Database queries monitoring & Results of database queries generated by application & Queries designated as cacheable by developers & Database triggers & Invalidation or Expiration-based \\ \hline

\citeN{Leszczynski2010} & Application and database monitoring & Database content & SQL statements & Dependency graph & Invalidation-based \\ \hline

\end{tabular}
\label{tab:static-consistency-approaches}
\end{table}

\citeN{Ports2010} introduced \emph{TxCache}, which consists of a caching approach with transaction support for data-driven web applications. TxCache associates cached content with invalidation tags, which indicate content dependencies. Every query processed in the database server, if not read-only, is returned along with its invalidation tags, which indicates content that is affected by the write operation (e.g.\ insertions, updates and removals). Write operations trigger invalidation processes that take the invalidation tags and evict all the related content in the cache, based on the presence of such tags. By doing so, TxCache provides a validity interval for cached content, instead of defining an arbitrary expiration time. Unlike TxCache, \emph{CacheGenie}~\cite{Gupta2011} keeps consistency by performing in-place updates rather than invalidating and recomputing data. CacheGenie is integrated with the object-relational mapping module provided by the web framework Django\footnote{\url{https://www.djangoproject.com/}} and takes as input caching configurations for each application object that is mapped to the database through a key-value structure. Such specification contains strategies, dependencies and other meta-data about caching and are used to automatically creating triggers within the database to evict and update cached data. Furthermore, CacheGenie also provides a weak consistency approach based on the definition of an expiration time, considering that many web applications do not require strong consistency approaches. 

Instead of requiring developer inputs, \citeN{Leszczynski2010} described an approach to cache data, which keeps it in a consistent state based on a dependency graph that provides a mapping between update statements in a relational database and cached content. When a database update is performed, the graph allows detection of cached content that must be invalidated to preserve the consistency of the cache and the data source. Also focused on keeping consistency by using the database as source of information, \citeN{Wang2014} and \citeN{Bouchenak2006} both propose solutions in the form of a middleware, which monitors database queries through JDBC and, based on the analysis of queries to track content dependencies, it triggers invalidations of all cached content that is related to a change, when a changing query is executed.

\subsubsection{Size-limitation Management}
\label{sec:static-sizelimitation}

Although we in this survey do not explore approaches that focus on general caching issues such as concurrency, scheduling, and other distributed systems topics, there are infrastructure-related issues that developers usually need to configure or at least be aware of, when implementing application-level caching~\cite{Mertz2016}, being cache size one of them. Because caches are size-limited, two main decisions must be made: (i) the definition of the adequate \emph{cache size}, and (ii) the choice of a \emph{replacement algorithm}. Table~\ref{tab:static-size-approaches} summarizes the caching approaches that address these problems of dealing with size limitation.

\begin{table}
\scriptsize
\centering
\renewcommand\tabcolsep{1mm}
\caption{Static Caching Approaches for Size-limitation Management.}%
\begin{tabular}{|p{2cm}|p{1.6cm}|p{2.6cm}|p{1.8cm}|p{2.9cm}|p{1.7cm}|}
\hline
\textbf{Approach} & \textbf{Based on} & \textbf{Data Input} & \textbf{Analysis} & \textbf{Output} & \textbf{Size-limitation Issue} \\ \hline

 Mimir \cite{Saemundsson2014} & Cache profiling & Cache size & Simulation-based & Hit ratio estimative according to the available cache size & Cache size{\par}definition \\ \hline

 \citeN{Zaidenberg2015} & Cache profiling & Recency and frequency of cached content & Heuristic-based & Cached content to be evicted & Replacement \\ \hline

 GD-Wheel \cite{Li2015} & Cache profiling & Recency and cost to retrieve of cached content & Heuristic-based & Cached content to be evicted & Replacement \\ \hline

\end{tabular}
\label{tab:static-size-approaches}
\end{table}

Regarding the first decision, \citeN{Saemundsson2014} proposed an insight into how much memory is necessary to trade-off the desired performance. Their approach is based on an online profiler, called \emph{Mimir}, which estimates how the popular least recently used (LRU) algorithm would perform with different ways of memory allocation, by continually exposing the hit ratio curve as a function of size cache. Thus, the approach allows \emph{what-if} questions, providing dynamic estimations of the cost and performance regarding different cache sizes.

While cache size is a key to improve the cache efficiency, replacement algorithms are fundamental as well. Such algorithms are usually based on simple heuristics such as recency, frequency or even randomly. Popular and well-accepted examples of replacement algorithms are the already mentioned LRU and least frequently used (LFU). Even though replacement seems a solved problem~\cite{Podlipnig2003}, new proposals have been developed. \citeN{Zaidenberg2015} presented a quantitative evaluation of alternative replacement strategies that consider both recency and frequency properties. They control data replacement through the management of different lists of cached items and levels of frequency and recency. The authors used LRU as a baseline and, according to benchmarks, the new proposals can increase the hit ratio. Despite this improvement, there are disadvantages, such as the sensibility to cache thrashing.

Recently, \citeN{Li2015} proposed \emph{GD-Wheel}, a cost-aware replacement policy based on the greedy-dual size (GDS) algorithm. GD-Wheel brings the exact GDS in the context of application-level caches, given that GDS was initially proposed by \citeN{Young1994} for web proxy caches. In a nutshell, besides the benefits of LRU, GDS also takes into account the cost to retrieve a piece of content while selecting content to be evicted~\cite{Cao1997}. Thus, GD-Wheel provides developers with a mechanism to include the cost information on each insertion or update in the cache. Such cost is taken into account in the replacement decisions. GDS works by maintaining a value of remaining priority $Rp$ for each cache entry. When inserting or reusing a piece of content $i$, $Rp(i)$ is defined as the cost of retrieving it $c(i)$. When an eviction is required, the piece of content with the lowest $Rp$ is evicted. After the eviction of $i$, all $Rp$ values of cached content are decreased by $Rp(i)$. By doing this, the algorithm integrates recency and cost to retrieve, because content with lower cost and not recently used has lower $Rp$ values, and consequently tends to be evicted faster.

A key limitation of the approaches presented in this section is that, due to the non-adaptive nature of these solutions, they do not automatically consider changes in application workload and access patterns, which can lead such approaches to a poor or sub-optimal performance before the revision of caching decisions. Furthermore, such approaches usually do not take application specificities into account and require human intervention for configuring, customizing and tuning the solution to the application requirements and needs.

\subsection{Adaptive Cache Management}
\label{sec:adaptiveapproaches}

We now focus on presenting the landscape of adaptive caching at application level and discuss research challenges in this context. Web caching approaches also include an underlying problem: the stochastic nature of user behavior, which may lead to unexpected or unpredicted workloads and this, consequently, affects the cache performance, especially in terms of coordination policies. This can cause, for example, cache thrashing, i.e.\ eviction of the useful data. Therefore, since their conception, it is desirable to make caching solutions conform to the dynamic changing of user demands and the network environment. Traditional cache strategies, such as replacement algorithms, already present a mechanism for adapting themselves over time, based on general heuristics like recency or frequency~\cite{Wang1999}. Although such strategies can somehow be seen as \emph{adaptive caching}, they are primarily conceived and implemented \emph{statically}. Therefore, exploiting concepts of adaptive systems could help design more robust, reliable, and scalable caching strategies.

In the remainder of this section, we discuss adaptive approaches for cache management, following the taxonomy shown in Figure~\ref{fig:approaches-taxonomy}. However, due to additional complexity of these approaches in comparison with static approaches, we first overview existing work and present alternatives used in the different activities of the feedback loop.

\subsubsection{Overview}

Adaptive approaches address different \emph{caching issues} and, for doing so, they \emph{analyze} a particular \emph{property} associated with caching. For example, if an approach focuses on consistency maintenance, it usually analyzes content changeability, because if the source of a cached content changes, it should be updated or removed from the cache. Based on this analysis, approaches adapt a particular task of cache management, e.g.\ definition of expiration time of cached item. This we refer to as \emph{adaptation target}. Broadly, adaptive caching approaches aim at increasing the cache performance as a means of improving the application performance. Selected measurements are used to evaluate the cache performance, and are thus associated with the \emph{expected outcomes} of a particular approach. We summarize the different adaptive caching approaches in Table~\ref{tab:adaptiveoptions}, analyzing these discussed approach characteristics as well as give examples from literature.

\begin{table}
\scriptsize
\centering
\renewcommand\tabcolsep{1mm}
\caption{Overview of Adaptation Scenarios for Application-Level Caching.}
\begin{tabular}{|p{1.7cm}|p{1.7cm}|p{2.7cm}|p{3cm}|p{3.8cm}|}
\hline
\textbf{Caching}\par\textbf{Issue} & \textbf{Analyzed}\par\textbf{Property} & \textbf{Adaptation Target} & \textbf{Expected}\par\textbf{Outcomes} & \textbf{Representative Examples} \\ \hline

  Content{\par}admission & Web content{\par}cacheability & -- Content selection{\par}-- Content filtering & -- Higher hit ratio & \cite{Baeza-Yate2007,Einziger2014,Chen2016}\\ \hline
  
  Consistency maintenance & Web content{\par}changeability & -- Expiration time{\par}definition & -- Higher hit ratio{\par}-- Increased content freshness & \cite{Huang2010}\\ \hline

  Content{\par}eviction & Web content{\par}evictability & -- Replacement decision{\par}-- Proactive eviction & -- Higher hit ratio{\par}-- Improved resource usage & \cite{AliAhmed2011,Ghandeharizadeh2014,Negrao2015}\\ \hline

  Cache{\par}infrastructure management & Cache resource{\par}availability & -- Cache size definition & -- Improved resource usage & \cite{Radhakrishnan2004}\\ \hline

\end{tabular}
\label{tab:adaptiveoptions}
\end{table}

Adaptation is usually driven by changes and behavior of the application environment and, therefore, \emph{monitoring} this environment is needed. This is typically done by capturing the execution traces of an application or cache to be analyzed, and deciding which actions to take. Such traces allow retrieving input data for the analysis, which can be mainly of two types. The first is \emph{stateless} data, which depend on the content that can be cached (e.g.\ the size of a query result, the frequency of a method call, and the recency of a requested content), and the second is \emph{stateful} data, which are based on historical information. The most straightforward approach is to use historical access information regarding the observed web component, such as requested URLs or database queries~\cite{Baeza-Yate2007,Ma2014,Chen2016}. Such monitored data can be easily obtained from web servers or database systems, which automatically store them in the form of weblogs. Moreover, the spatial locality can be used by analyzing the content context~\cite{Negrao2015}. Although such contextual information can be useful, it requires a deeper analysis of the content being managed.

The analysis of monitored data requires the characterization of the execution of an application or cache. Such task can be done based on simple heuristics~\cite{Negrao2015}, or by using sophisticated techniques, such as a mathematical model~\cite{Huang2010,Chen2016} and machine learning~\cite{Sajeev2010b}. Furthermore, static analysis of source code can also be useful~\cite{Chen2016}. Such static analysis can reduce the complexity and processing required to achieve the caching decisions, i.e.\ reduce the \emph{learning phase}. For example, while looking for cacheable opportunities, static analysis can provide a limited set of possibilities and places to look at, thus reducing the amount of data that should be analyzed, making the decision process faster and more efficient.

The decision can be implemented also with heuristics~\cite{Negrao2015}, utility-functions~\cite{Baeza-Yate2007,Ma2014,Einziger2014}, or even taking into account the output of a classifier~\cite{Sajeev2010b}. Finally, the goals can be achieved by effectors connected directly with the cache~\cite{Ma2014,Einziger2014}, the application~\cite{Chen2016}, or even with a middleware component.

The general activities presented and exemplified above (i.e.\ monitoring, analyzing, deciding and adapting) are seen in self-adaptive systems as a feedback loop mechanism. Based on such a loop, these systems are able to change their behavior in response to noted changes in their environment. A deeper discussion about autonomic computing and self-adaptive systems is given by \citeN{Huebscher2008} and \citeN{Lalanda2013}. In this survey, we capture characteristics of existing approaches matching the activities of a feedback loop to facilitate understanding and comparison of the approaches, even if an approach is not explicitly structured in this way. A feedback loop of an adaptive system can be described in terms of six key properties, as described below. These properties are used to compare and describe adaptive approaches in later sections.

\begin{description}
\item[Monitored data] Monitored data are measurable input parameters from which the current state of the system can be characterized.

\item[Analysis] Based on the monitored data, an analysis process should be employed to characterize the state of the system. Such characterization can be based on a single monitored piece of information or a combination of a set of inputs, such as a utility function, or even a complex model built through machine learning approaches.

\item[Behavior] The behavior represents how the adaptive component acts over the managed resource or system. It can be reactive, i.e.\ responding to observed events after it occurs, or proactive, i.e.\ taking actions in advance before events have a chance to happen.

\item[Operation] The operation corresponds to when and how the adaptation process can analyze data and take actions. Online operation is the case when the adaptation process takes place while the application is executing, possibly impacting in its functioning. The adaptation process can be seen as part of the application. Offline operation, in contrast, is performed separately, based on the information collected from the application to a later analysis. Actions to be taken are performed when deemed adequate.

\item[Decision] Decisions are conclusions or resolutions achieved after the analysis. Such decisions involve reasoning about the current state of the system and defining necessary adaptations (if needed) to achieve the goals.

\item[Goal] Goals consist of one or more expected properties that should be achieved by the adaptations. They summarize the desired system performance or behavior regarding input parameters.
\end{description}

\subsubsection{Content Admission}

We now focus on approaches that can dynamically evaluate the cacheability status towards discovering cacheable data, i.e.\ whether a particular piece of content should be cached. As previously mentioned, admission approaches help developers while selecting caching opportunities. This task is particularly complex because selected opportunities must continuously be revised, due to the changing workload characteristics and access patterns, or even the application evolution. This shortcoming motivates the need for adaptive caching solutions, which can minimize the effort required from application developers to maintain caching solutions by automatically improving themselves regarding changes in the application context.

As shown in Figure~\ref{fig:approaches-taxonomy}, adaptive admission-focused approaches can be seen from two different perspectives, depending on the purpose of the adaptation: (a) \emph{dynamic selection} of the best cacheable opportunities through the evaluation of cacheability properties, and (b) \emph{reactive filtering} of content that was previously identified as cacheable but should not anymore, due to some particular reason. Table~\ref{tab:adaptive-admission-approaches} summarizes the surveyed adaptive caching approaches that deal with admission issues.

\begin{table}
\scriptsize
\centering
\renewcommand\tabcolsep{1mm}
\caption{Comparison of Adaptive Application-level Caching Approaches for Admission.}
\begin{tabular}{|p{1.7cm}|p{2cm}|p{2cm}|p{1cm}|p{1.15cm}|p{2.2cm}|p{2.3cm}|}
\hline
\textbf{Reference} & \textbf{Monitored Data} & \textbf{Analysis} & \textbf{Behavior} & \textbf{Operation} & \textbf{Decision} & \textbf{Goal} \\ \hline

  CacheOptimizer \cite{Chen2016} & Web server and database access logs & Static source code analysis and dynamic weblog analysis with colored Petri nets & Proactive & Offline & Content miss ratio threshold & Dynamic identification of cacheable content and definition of caching configurations \\ \hline

  \citeN{Mertz2017,Mertz2017b} & Method calls & Cacheability metrics & Proactive & Online and offline phases & Heuristic-based & Dynamic identification and caching of cacheable methods \\ \hline

  \citeN{Baeza-Yate2007} & Query metadata and historic usage information & Utility-function based & Reactive & Online & Utility-function based on the probability of a query generate future cache hits & Dynamic decision regarding the admission of queries to the cache \\ \hline

  TinyLFU \cite{Einziger2014} & Cache size (number of items it can store) and historic usage information & Frequency histogram & Reactive & Online & Utility-function based on the potential hit ratio increasing & Dynamic decision regarding the admission of content when an eviction is required \\ \hline

  LWE-LRU \cite{Sajeev2010b} & Content attributes and traffic parameters & Multinomial logistic regression model to compute worthiness of content & Reactive & Offline and online phase & Worthiness thresholds & Dynamic admission and eviction decisions \\ \hline

\end{tabular}
\label{tab:adaptive-admission-approaches}
\end{table}

The automatic identification of caching opportunities at the application level is addressed by \emph{CacheOptimizer}~\cite{Chen2016}, which monitors readable weblogs to create mappings between workload and database access. The analysis part consists of two processes: a static code analysis to identify possible caching configuration spots, and a characterization with colored Petri nets, which models the transition of states of an application based on weblogs from the web server. By doing so, it can reach a global optimal caching decision, instead of focusing on top cache accesses, using a greedy approach. Differently from other approaches, CacheOptimizer is not implemented as a new caching framework; instead, it is integrated with those existing, automating the caching configuration according to the results of its analysis. Although this approach addresses method caching opportunities, it focuses on database-centric web applications; thus, only database-related methods are cached. By focusing on general application methods as cacheable options, \citeN{Mertz2017,Mertz2017b} proposed a seamless approach that automates the evaluation of cacheability properties and admits application methods at runtime based on the Cacheability Pattern~\cite{Mertz2016}. These approaches were evaluated considering single-shot scenarios, without having their adaptive behavior measured.

Reactive filtering approaches act at runtime by dynamically evaluating the cacheability of content that was previously identified as cacheable, according to the workload and access pattern, to avoid unworthy content taking place in the cache. By doing this, it can reduce the amount of cached data, freeing space in the cache, avoiding evictions and leading to higher hit ratios. Reactive filtering approaches are specifically useful when dealing with search engines, since not all search combinations will frequently be requested as shown by \citeN{Baeza-Yate2007}. They proposed an approach to filter infrequent searches, which would not improve application performance if cached. The proposed admission solution monitors query executions and caches such queries into a split cache scheme: controlled and uncontrolled. Both parts internally implement a regular LRU-based cache, but an admission policy is used to distinguish frequent queries, which are placed into the controlled space, from infrequent queries, which are placed into the uncontrolled space. The admission policy is implemented as a function that evaluates each processed search query taking into consideration the query metadata and past usage information. As result, the controlled cache is less pollution-prone, and frequent queries tend to remain cached for longer periods, improving hit ratios. Although infrequent queries are more rapidly evicted in the uncontrolled space, it is still an LRU-based cache and can provide good responses in cases of short-period bursts of infrequent queries.

Also focusing on reactively filtering content, \citeN{Einziger2014} proposed \emph{TinyLFU}. Despite such approach acts only when the cache is full, it admits new content in the cache only when such content is expected to provide more benefit than the content already in the cache. TinyLFU estimates the worthiness (i.e.\ contribution to the hit ratio) of the eviction candidate and compares it with the worthiness of the newly accessed content. To achieve such behavior, TinyLFU uses an approximate LFU structure, maintaining a frequency histogram for cached content, and then it is possible to trade-off the cost of eviction and the usefulness of the new content. A small variation of such approach is currently available to developers as a default replacement policy of the Caffeine caching framework, due to its high hit rate and low memory footprint.

Admission approaches can also be implemented by using sophisticated learning techniques. \citeN{Sajeev2010b} proposed an admission technique using multinomial logistic regression (MLR) as a classifier. The model built with MLR assigns a worthiness class to the response of a processed request based on the application traffic and response content properties. Such model is trained with previously collected traces and sanitized logs for classifying the web cache's content worthiness. Then, at runtime, when there is an incoming content, its worthiness class is computed and used in an admission control mechanism based on thresholds to decide whether the content should be cached or not.

\subsubsection{Consistency Maintenance}
\label{subsec:adaptive:consistency}

Adaptation in consistency maintenance can be employed in both introduced consistency approaches, i.e.\ expiration-based and invalidation-based. However, differently from static approaches, existing adaptive work focuses only on the former, as can be seen in Figure~\ref{fig:approaches-taxonomy}. Consequently, all the approaches assume that stale data is allowed to be returned from the cache, that is, they are based on weak consistency. Table~\ref{tab:adaptive-consistency-approaches} summarizes the surveyed adaptive caching approaches that deal with consistency issues.

\begin{table}
\scriptsize
\centering
\renewcommand\tabcolsep{1mm}
\caption{Comparison of Adaptive Application-level Caching Approaches for Consistency.}
\begin{tabular}{|p{1.8cm}|p{2.3cm}|p{1.6cm}|p{1cm}|p{1.15cm}|p{2.2cm}|p{2.3cm}|}
\hline
\textbf{Reference} & \textbf{Monitored Data} & \textbf{Analysis} & \textbf{Behavior} & \textbf{Operation} & \textbf{Decision} & \textbf{Goal} \\ \hline

  \citeN{Huang2010} & Hit ratio and historic usage information & Function approximations & Reactive & Online & Utility-function based on the monitored data & Dynamic definition of the expiration time of data \\ \hline

  Adaptive, Incremental and Machine-learned TTL \cite{Alici2012} & Query metadata and historic usage information & Metric-based & Reactive & Online & Rule-based & Dynamic definition of the expiration time of data \\ \hline

\end{tabular}
\label{tab:adaptive-consistency-approaches}
\end{table}

In such approaches, it is assumed that cached content is accessed irregularly, and then the definition of a fixed timeout is inadequate. If the cached content has a short expiration time, it tends to lower hit ratios in situations different from a short burst of requests. Otherwise, if it has a long expiration time, it can result in stale data being returned to users. Despite staleness is expected in weak consistency approaches, the level in which it affects the system execution should be taken into account by developers. Due to this, \citeN{Huang2010} proposed a cache framework that provides transparency for developers by offering an adaptive expiration time---also called time-to-live (TTL)---definition. The proposed strategy is based on the rationale behind the popular simulated annealing algorithm, in which an expiration time function is approximated according to changes into how users access content. Also focused on TTL definition, \citeN{Alici2012} proposed a query result caching for search engines, which considers query-specific TTL values, using a machine learning model to assign TTL values to queries. Their machine learning model exploits a query log and result-specific features to predict the best TTL values. The authors showed that this strategy outperforms the fixed TTL strategy, by means of an evaluation of some parameterized functions (average and incremental TTL). 

\subsubsection{Size-limitation Management}

Cache frameworks and libraries are usually based on a fixed memory size to allocate content. It can be done in terms of allowed number of entries or absolute size. As introduced, there are static approaches that help determine the appropriate cache size and deal with replacement issues that emerge due to a size-limited cache. There are also adaptive approaches that support them, presented and compared in Table~\ref{tab:adaptive-replacement-approaches}.

\begin{table}
\scriptsize
\centering
\renewcommand\tabcolsep{1mm}
\caption{Comparison of Adaptive Application-level Caching Approaches for Dealing with Size Limitations.}
\begin{tabular}{|p{1.7cm}|p{2cm}|p{2cm}|p{1cm}|p{1.15cm}|p{2.2cm}|p{2.3cm}|}
\hline
\textbf{Reference} & \textbf{Monitored Data} & \textbf{Analysis} & \textbf{Behavior} & \textbf{Operation} & \textbf{Decision} & \textbf{Goal} \\ \hline

  \citeN{Radhakrishnan2004} & Hit ratio and cache size & Linear quadratic estimation & Proactive & Online & Based on pre-defined thresholds & Dynamic increasing or decreasing cache size \\ \hline

  ARC \cite{NimrodMegiddo2003} \par ARC-TD \cite{Raigoza2014} & Recency and frequency of requests to cached data & Heuristic-based & Reactive & Online & Likelihood of content being used in the near future & Dynamic selection between recency-based or frequency-based eviction \\ \hline

  \citeN{ElAbdouniKhayari2011} & Cache access logs and metadata, such as requested data sizes, arriving times, hits and misses & 
  Expectation Maximization (EM) algorithm & Reactive & Offline and online phase & Output of EM algorithm & Dynamic reconfiguration of C-LRU algorithm \\ \hline

  CAMP \cite{Ghandeharizadeh2014} & Cost to retrieve, size and recency of cached web content & Heuristics based on cost-to-size ratio and recency & Reactive & Online & Based on priority value of content & Dynamic selection of the best candidate to evict \\ \hline

  SACS \cite{Negrao2015} & Web content along with temporal information and pre-defined thresholds & Metric of distance between content  & Reactive & Offline and online phase & Higher distance & Dynamic selection the best candidate to evict \\ \hline

  ICWCS \cite{AliAhmed2011} & Web access logs & Adaptive neuro-fuzzy inference system & Proactive & Offline & Trained neuro-fuzzy system that models the access probability of web content & Dynamic eviction of unworthy cached web content \\ \hline

\end{tabular}
\label{tab:adaptive-replacement-approaches}
\end{table}

\citeN{Radhakrishnan2004} addressed the trade-off between cache space allocation and performance improvement by dynamically changing the cache size according to the access patterns and workload. It uses a linear quadratic estimation (i.e.\ kalman filtering) algorithm that estimates properly cache size values based on runtime information (hit ratio and memory in use) collected over time and pre-defined thresholds. Given the cache size estimation, the approach is able to decide which action promotes more benefits: increasing or decreasing the cache size, or even keeping it as it is and replacing content.

Regarding replacement policies, popular algorithms such as LRU and LFU have the limitation of considering a single factor, possibly ignoring important properties that can influence the replacement efficiency. In realistic applications, access patterns significantly change over time, and simple replacement policies may not have enough information to properly deal with these situations. A common problem is \emph{cache pollution}, in which the cache is loaded with unnecessary data, resulting in useful data being evicted. Such problem can be either cold (affecting LRU) or hot cache pollution (affecting LFU and size-based policies)~\cite{Ayani2003}.

As shown in Figure~\ref{fig:approaches-taxonomy}, adaptive replacement approaches can be classified into three groups: (a) \emph{dynamic selection of algorithms}, which focuses on exploring the advantages of each algorithm, (b) \emph{evolution of traditional replacement policies} to achieve adaptation, including the modification of replacement approaches proposed originally for other caching levels (e.g.\ web proxy caching and database), and (c) the proposal of new and \emph{application-level specific} approaches.

One of the first initiatives to adapt replacement policies is the \emph{Adaptive Replacement Cache (ARC)}~\cite{NimrodMegiddo2003} approach, which combines the merits of different replacement policies, and dynamically balances between the recency and frequency components online. The policy keeps: (i) the cache space is partitioned according to defined constraints in order to hold content that was accessed recently and frequently (at least twice); and (ii) a recent eviction history of the partitions. It uses a learning rule to continually revise the adaptation parameter in response to the observed workload. A practical implementation done by the same authors~\cite{Megiddo2004} revealed a better hit ratio with ARC over LRU across a wide range of workloads while incurring practically the same low time cost as the LRU. Although it became popular and is currently implemented in frameworks available to developers, such as Caffeine, it is patented\footnote{\url{https://www.google.com/patents/US20040098541}} and requires a license agreement with IBM to be used. Furthermore, \citeN{Raigoza2014} proposed the \emph{Adaptive Replacement Cache-Temporal Data (ARC-TD)}, in which a buffer replacement policy is built upon the ARC policy. Both ARC and ARC-TD policies address the problem of caching priority by trading-off between frequently accessed and recently used content. However, the ARC-TD policy favors the cache retention of content in proportion to the average life span of the content in the buffer. Thus, a higher cache hit ratio can be achieved.

\citeN{ElAbdouniKhayari2011} presented a runtime closed-loop for self-reconfiguration of the \emph{Class-Based Least Recently Used (C-LRU)} strategy. In C-LRU, the cache is divided into portions reserved for content of a specific size, and the content size distribution is described by a hyper-exponential distribution using the EM-algorithm. The approach is based on a continuous analysis of the cache trace (access logs), from which relevant data, such as the requested data sizes, the arriving times, hits and misses, are retrieved. Then, it computes the parameters for a hyper-exponential distribution function of the request content sizes, which are used to determine a new configuration for the C-LRU.

Recently, by using cost as an important feature to guide replacement decisions, \citeN{Ghandeharizadeh2014} proposed the \emph{Cost Adaptive Multi-Queue Eviction Policy (CAMP)}, which is an adaptation of the GDS algorithm~\cite{Young1994} (as described in Section~\ref{sec:staticapproaches}). CAMP manages multiple LRU queues based on the distribution of cost and size of cached content. Each queue is ordered according to the remaining priority of the items. When a replacement is required, CAMP finds the eviction candidate across all queues based on looking the remaining priority at the front of each queue, which conveys the item with the smallest priority. Such approach also has a variation in which FIFO queues are adopted~\cite{Ghandeharizadeh2015}, because it demands fewer content moves.

As shown, most of the adaptive replacement approaches use temporal locality information to make replacement decisions. However, a recent work done by \citeN{Negrao2015} proposed the consideration of spatial locality of cached content while finding an eviction candidate. \emph{Semantics-Aware Caching System (SACS)} incorporates the assumption that content (in this case, web pages) that can be achieved from recently requested content tends to be requested shortly and, thus, should also be kept in the cache. Such assumption is captured by a distance metric, which is calculated as the minimum number of links that users are supposed to follow to navigate from one page to the other. Then, while selecting content to evict, SACS takes the most recently requested content (given a threshold) as pivots and calculates the distance of other cached content from pivots. Older cached content with higher distance values is removed until there is enough space to insert the new content. Thus, cached content that is spatially near to recently requested content is kept cached. In addition, when the same distance occurs in two or more entries, the frequency is used as the next criterion to make the decision.

Finally, focused on the client-side, \citeN{AliAhmed2011} proposed a neuro-fuzzy system, called \emph{Intelligent Client-side Web Caching Scheme (ICWCS)}, that estimates content that can be accessed in the future. This approach splits the cache space into short-term and long-term caches, in which the former is an LRU-based cache that receives content directly, and the latter receives content evicted from the short-term cache. When the long-term cache achieves its full capacity, a trained neuro-fuzzy system is employed to distinguish content in the long-term cache as cacheable or uncacheable. Then, content classified as uncacheable are evicted.

\section{Conclusion and Future Directions}
\label{sec:conclusion}

With the significant amount of users that web applications currently serve, meeting performance and scalability requirements while delivering services has been a crucial challenging issue. To address such requirements, various web caching technologies were made available in different layers of the web infrastructure. Recently, an application-tailored form of caching has increased in popularity, referred to as application-level caching, in which developers manually insert caching logic into the application base code to decrease the response time of web requests by temporarily saving frequently requested or expensive to compute content in memory. However, such caching leads to many issues and challenges for developers, concerning the design, implementation, configuration, tuning, and maintenance.

In this paper, we provided a comprehensive introduction to web and application-level caching, as well as surveyed approaches proposed in the context of the latter. We first presented a taxonomy of web caching in general, which helps understand the different associated concerns as well as the drawbacks and merits of available web caching options. We also provided an overview of application-level caching, highlighting its challenges and issues, followed by a big picture of the different approaches proposed to support its development. Application-level caching approaches were described and compared, grouped into three categories: cache implementation, static cache management, and adaptive cache management.

Our survey provides insights into the theory of application-level caching solutions and brings important points to the attention of researchers and practitioners. The survey can be used by developers to learn existing caching solutions available so that they can improve their web applications, and by researchers to have a solid foundation on this topic for developing future adaptive caching approaches. We envision that such approaches will require minimal effort and input from developers, as opposed to addressing assumed application bottlenecks through simple key-value systems that require manual management and tuning. We next wrap up our survey by providing reflections on the field and discussing future challenges to be addressed.

\subsection{Reflections on the Field}

Caching techniques can be found in many locations of computer and software architectures and, in all of them, such solutions provide benefits, considering the assumption that part of the content is repeatedly accessed, which is often valid. Web caching, in general, has been extensively studied in the last decade, especially for web pages (at the proxy level), and many issues have been brought up, discussed and addressed. Although web caching in its general form seems to be a resolved topic, there are open gaps that must be filled, regarding particular systems and requirements, such as multimedia (with many large objects), wireless (with higher latency and error rate), mobile systems and Internet of things (with mobility and smaller cache), and modern web applications (with customized content and higher changeability), being the last the focus of this survey.

In these specific caching scenarios, the performance benefits of a caching solution highly depend on exploiting the peculiarities of each scenario. However, while proposing new caching solutions, it is fundamental to learn from different areas. First, we must learn and leverage what has been done in traditional and older forms of caching, such as in computer architectures. Second, we must complement information about a particular domain with user access patterns, as well as computer and network characteristics. Third, to cope with the evolution of the application and its environment, self-adaptive techniques can be exploited. As discussed, this last point could be much further explored. Although it is possible to interpret many adaptive approaches in terms of the traditional activities of self-adaptive systems, they lack an explicit connection with work on this area. This hampers the exploitation of existing self-adaptive techniques in our context.

The continuously changing web environment is the main driver for the adoption of adaptive web caching approaches, but the availability of web access logs and the existence of specific tools for constantly monitoring resources (application and cache execution traces) have also motivated the development of adaptive techniques. As a consequence, the runtime environment can be instrumented to gather useful information to support dynamic changes in the system behavior, adjust in misconfiguration, or even achieve optimal configurations. Approaches that monitor the application execution and change its inner workings tend to perform better than a static and pre-defined solution~\cite{Ali2011}.

Even though there are approaches that focus on providing application-level caching with an adaptive behavior, only a few consider application specificities. Usually, caching decisions are made based solely on access information (size, recency, and frequency) and statistics (hit and miss ratios), thus ignoring expressed cache metadata, which is application-specific. Application details are in fact information that developers use while designing and implementing application-level caching. Therefore, the discovery of application-specific details is crucial to achieve an improved performance of caching.

An important step in the direction of such approaches is to understand and structure knowledge regarding application-level caching spread in existing caching solutions for web applications. There is a qualitative study~\cite{Mertz2016} in this direction that investigated how developers approach application-level caching through the analysis of ten (open-source and commercial) web applications. Furthermore, other works~\cite{Atikoglu2012,Nishtala2013,Xu2014} provided an analysis of the workload of application-level caches, giving characterizations and a detailed picture of these components. This kind of work is needed to not only have a solid understanding of how such caches are currently used but also for giving future directions to improve this form of caching.

\subsection{Major Challenges and Research Issues}

Providing an adaptive approach requires addressing different issues, highlighted in the feedback loop of self-adaptive systems. However, while designing and implementing ways to collect, analyze and make decisions regarding adaptation at application-level, other specific issues emerge.

The first issue is associated with the identification of what information to collect as part of the monitoring process. Despite web access logs are the most common source of information, because it provides the history of access to a given resource, a previous analysis of the application regarding static information, such as the application source code, is also useful to understand the behavior of the application~\cite{DellaToffola2015,Chen2016}. The combination of static and dynamic analyses can potentially improve adaptation decisions as they are complementary, or even reduce the information collected at runtime required by the dynamic analysis. In addition, it can lead to better initial decisions, when limited information is available.

The second issue is how to adequately monitor and acquire information from the application and its environment without compromising its performance. Such monitoring should be as non-intrusive as possible to avoid affecting the performance of the application or cache, which is not always possible~\cite{Chen2016}. Adaptive approaches should be designed considering the trade-off between the gain of automating caching concerns and the application performance loss due to adaptation overhead. In fact, even a relatively simple analysis of the application profiling required by adaptive approaches is already responsible for a possibly high overhead regarding memory consumption and processing time. The overhead of the monitoring activity was not explored in the surveyed approaches, despite mentioned by some of them~\cite{Mertz2017,Mertz2017b}. Consequently, it is important to evaluate the practical feasibility of integrating such approaches at runtime with web applications, in terms of the impact they cause to collect data. Moreover, alternatives to reduce the impact of the monitoring activity should be explored.

Another aspect that deserves attention is the consideration of uncertainty in estimating application demands while evaluating application-level caching solutions. Configurations based on misleading information can result in a performance decay and increase the response time of the application. Furthermore, the adoption of standard benchmarks for evaluating and fine-tuning mechanisms of self-adaptation would help researchers and practitioners assess the performance and suitability of different caching algorithms.

Regarding caching algorithms, replacement policies are the most studied techniques because it is a straightforward solution for the cache fulfillment problem~\cite{Podlipnig2003}. However, size limitations can be solved from other perspectives, such as by designing a better admission policy~\cite{Baeza-Yate2007,Einziger2014}. Furthermore, the main issue regarding the development of consistency and admission approaches is that such techniques are highly dependent on the current application characteristics, as well as the required consistency level. Consequently, providing means of learning these is a key to the provision of satisfactory caching solutions.

In addition to consistency and admission policies, prediction-based admission (prefetching), which focuses on anticipating the cache of useful content, has also been proposed in other web caching levels, mainly for web proxy caches~\cite{Ali2011}. This topic has not been explored at the application level yet, given that it usually requires a complex learning phase to guide the selection of future content. Moreover, spatial locality---an important feature to be analyzed while prefetching content because it gives the notion of nearness---is not trivial to be used, due to the difficulty in identifying and collecting information that represents this property.

Last, but not least, a feature provided by application-level caching approaches is the possibility for developers to configure and tune them~\cite{Guo2011,Gupta2011}. Therefore, it is important to give to application developers the possibility of enriching the application model at design time with valuable application-specific information regarding the applicability of caching. This metadata can be processed and used by approaches, easing the process of analyzing and achieving adaptations. Moreover, this opens the possibility of caching approaches to use application-specific knowledge, which is the key reason why application-level caching exists. However, many issues must be addressed to capture such knowledge, such as understanding what sort of information must be provided,  how they are modeled, and how to keep them updated.

In summary, despite significant advances have been made in the context of application-level caching and adaptive caching at web applications, research on such topics still offers several opportunities for future research. The presented issues and shortcomings, especially regarding adaptive solutions, call for new approaches and techniques, to reduce the effort demanded from developers to design, implement, and maintain web caching solutions.

\begin{acks}
We thank the anonymous reviewers who were asked by ACM Computing Surveys to review this article, as well as the editor. They provided constructive feedback that was used to expand our research and improve this article extensively. Jhonny Mertz would like to thank CAPES grant ref. 1681815. Ingrid Nunes thanks for research grants CNPq ref. 303232/2015-3, CAPES ref. 7619-15-4, and Alexander von Humboldt, ref. BRA 1184533 HFSTCAPES-P.
\end{acks}

\bibliographystyle{ACM-Reference-Format}
\bibliography{references}

\begin{appendices}
\section{Background on Caching and Web Caching}
\label{sec:appendix:background}

In this section, we provide foundations on caching and web caching, presenting definitions and explanations of associated terms and concepts. This gives the basic but required knowledge for understanding the static and adaptive application-level approaches presented in the paper.

\begin{figure}
  \includegraphics[width=\textwidth]{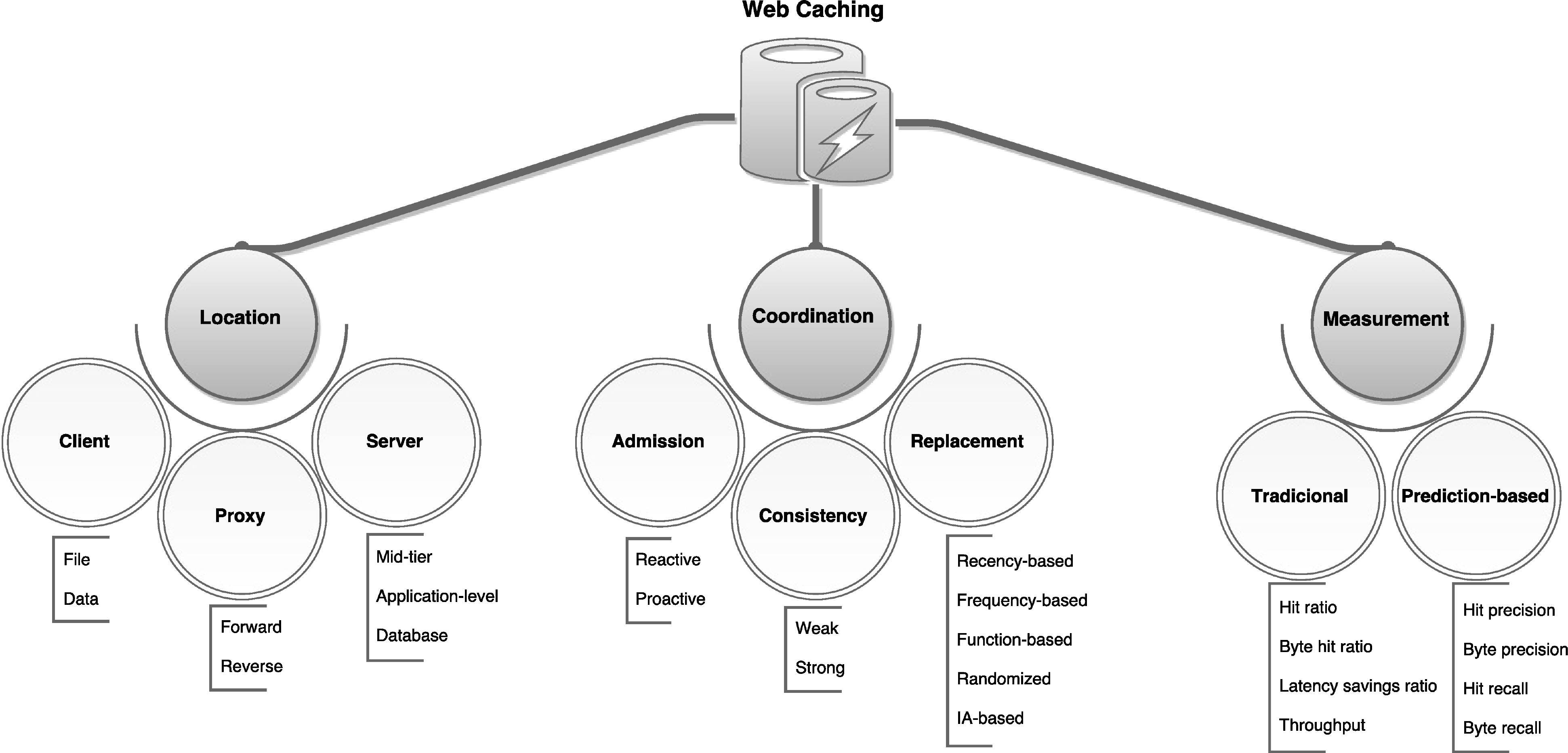}
  \caption{Web Caching Taxonomy.}
  \label{fig:cachetaxonomy}
\end{figure}

To introduce web caching, we use the taxonomy presented in Figure~\ref{fig:cachetaxonomy}, which groups web caching research into three main categories, each split into subcategories. The first category, \emph{location}, is associated with where caching is located in the typically used web architecture. The second is concerned with how the caching is \emph{coordinated}, which involves different tasks. Finally, the last is associated with how caching can be \emph{measured}, to have its effectiveness assessed. We next explain each of these categories and their subcategories. Every caching solution can be classified according to these three categories, given that its development involves choosing a location, dealing with coordination issues, and selecting appropriate metrics to estimate the performance of the provided solution.

\subsection{Location}
\label{subsec:architecture}

\begin{figure}
  \centering
  \includegraphics[width=\textwidth]{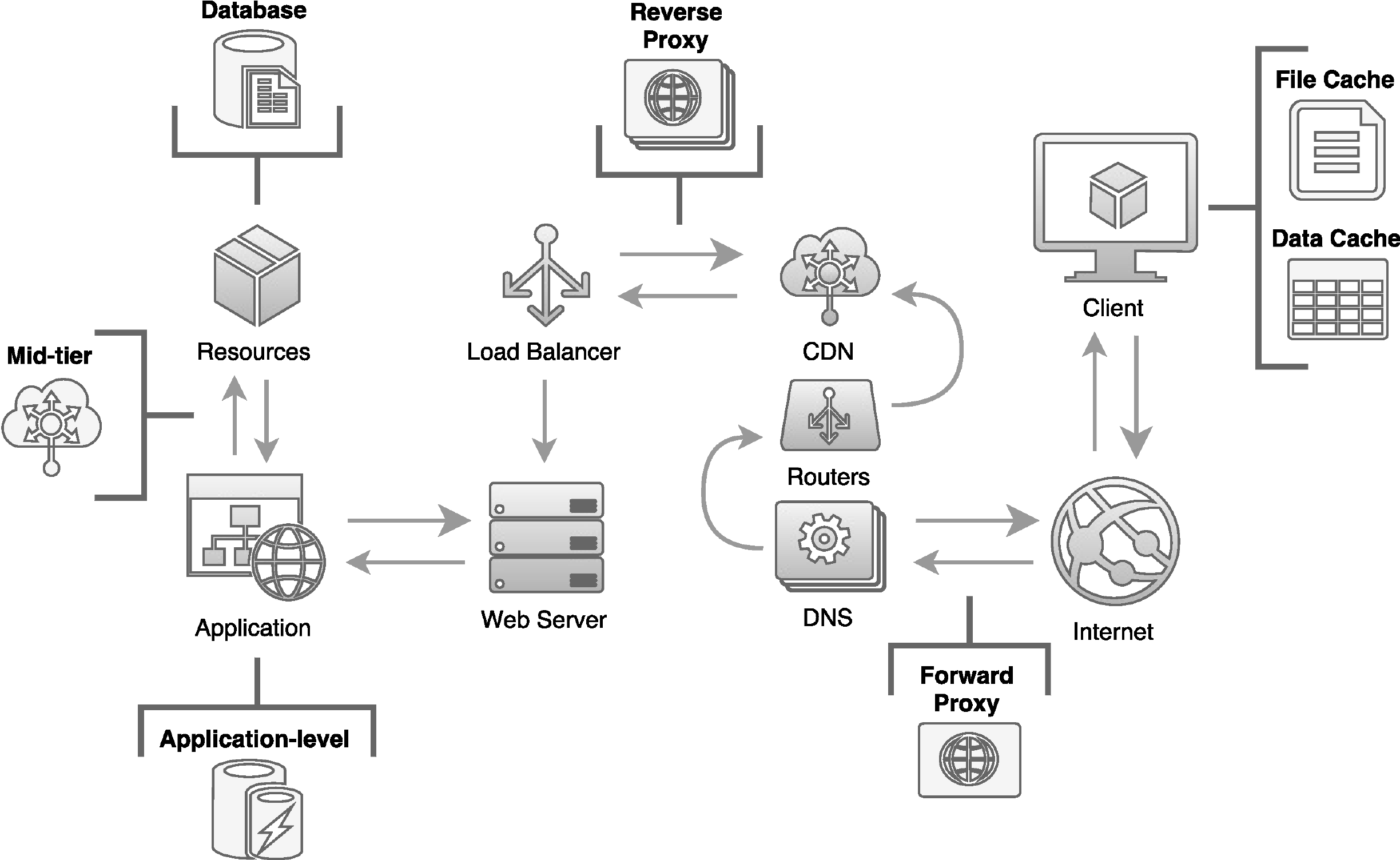}
  \caption{Traditional Web Application Architecture with Associated Caching Locations.}
  \label{fig:webinfrastructure}
\end{figure}

Given that web caching is closely related to the web infrastructure components, we overview such architecture in Figure~\ref{fig:webinfrastructure} along with the main caching locations, which are detailed later. It presents a typical web architecture, which is widely adopted in practice and can roughly be separated into three components: client, Internet, and server~\cite{Labrinidis2009,Ravi2009}.

The \emph{client} component is essentially the client's computer and web browser; while the \emph{Internet} component contains a wide range of different, interconnected mechanisms to enable the connection between client and server. A \emph{Domain Name Server} is typically used to decode the name of the web address into an Internet Protocol (IP) address. With this IP address, routers are used so that the client can establish a connection with the server, using the HTTP protocol. The \emph{server} can include multiple and different servers that are collectively seen as the web server by the client. In this case, a load balancer or similar component is used to route requests to the different servers in the cluster. A single web server is responsible for handling and responding to HTTP requests, and typically hosts a web application, which provides business services. The web application is connected to back-end systems (i.e.\ databases, file systems or other applications) and performs read and write operations driven by business rules~\cite{Labrinidis2009,Ravi2009}. In summary, caching solutions can be deployed at different locations along the typical web architecture, highlighted (in bold and inside brackets) in Figure~\ref{fig:webinfrastructure}, ranging from the database to the client's browser~\cite{Podlipnig2003}. We classify web caching locations into three main locations, as follows.

Caching solutions can be deployed at different locations along the typical web architecture, ranging from the database to the client's browser~\cite{Podlipnig2003}. We classify web caching locations into three main locations, as follows.

\begin{description}

\item[Server] Server-side caching solutions include all available caching locations within the server component, such as in databases, between database and application, within the application, and in the web server~\cite{Ravi2009}.

\item[Proxy] Located between client and server, a proxy caching can be employed in a forward (proxy on behalf of clients) or reverse (proxy on behalf of servers) way~\cite{Podlipnig2003,Ravi2009}.

\item[Client] At the client level, browser data, and client-side computations can be cached near the client, more specifically, on a machine where the users' web browser is located~\cite{Wang1999,Podlipnig2003,Balamash2004}.

\end{description}

Usually, caching solutions are designed and implemented as an abstraction layer that is transparent from the perspective of adjacent layers. Database and proxy caching are well-known for following this design principle. Differently, application-specific caches, which can be conceived at the server-side or client-side, require an extra effort from developers to be integrated into the web infrastructure~\cite{Gupta2011,Wang2014,Ports2010}. Given these several available caching locations and their pros and cons, we summarize this information in Table~\ref{tab:cacheclassification}.

Each caching location has its benefits, challenges, and issues, leading to trade-offs to be resolved when choosing a caching solution. For instance, according to a selected choice of location, particular forms of content can be cached, such as HTML pages~\cite{Candan2001,Negrao2015}, intermediate HTML or XML fragments~\cite{Ramaswamy2005,Li2006a,Guerrero2011}, application objects~\cite{Ports2010,Gupta2011}, database queries~\cite{Soundararajan2005,Amza2005,Baeza-Yate2007,Ma2014}, or database tables~\cite{Larson2004}. Furthermore, support varies across different locations~\cite{Mehrotra2010}, as well as hit and miss probabilities. Therefore, it is important to make an effort to achieve the best design rationale to optimize each caching solution according to different circumstances. Given these several available caching locations and their pros and cons, we summarize this information in Table~\ref{tab:cacheclassification}.

\begin{table}
\renewcommand\tabcolsep{1mm}
\scriptsize
\caption{Available Web Caching Techniques based on the Location.}%
\begin{tabular}{|c|p{1.2cm}|p{2.4cm}|p{5.3cm}|p{3.7cm}|}
\hline
  & \textbf{Location} & \textbf{Content \newline Granularity} & \textbf{Positive Aspects} & \textbf{Negative Aspects} \\ \hline

  \multirow{2}{*}[-6ex]{\rotatebox[origin=c]{90}{Client-side}}
    & File & -- Static resources \par-- Pages or fragments & + Integrated into the client browser \par + No development effort required \par + Effective for static content \par + A hit provides a large gain \par + Larger reductions in perceived latency & -- Harder to maintain consistency \par -- Cached content cannot be shared \par-- Manually configured at server-side \par-- Lower hit ratio \\ \cline{2-5} 
    & Data & -- Requests from a single client to many servers \par -- Computations & + Larger reductions in perceived latency \par+ Addresses arbitrary content \par+ Flexible implementation \par+ Takes application specificities into account & -- Manually implemented \par-- Cached content cannot be shared \\ \hline 

  \multirow{2}{*}[-3ex]{\rotatebox[origin=c]{90}{Proxy-based}} 
    & Forward & -- Requests from many clients to many servers in the gateway server & + Wide area bandwidth savings and improved latency \par+ Increases the availability of static content \par+ Fully transparent to developers & -- Harder to maintain consistency \par-- Requires a large-sized memory \\ \cline{2-5} 
    & Reverse & -- Requests from many clients to one server & + Reduces bandwidth requirements \par+ Reduces server-side workload \par+ Fully transparent to developers \par+ Easier to set up & -- Harder to maintain consistency \par-- Requires a large-sized memory \par-- Lower hit ratio \\ \hline
 
  \multirow{3}{*}[-14.5ex]{\rotatebox[origin=c]{90}{Server-side}}
    & Application & -- Pages or fragments \par-- Service calls \par-- Database queries \par-- Computations & + Addresses arbitrary content \par+ Easier to maintain consistency \par+ Saves processing requests \par+ Flexible implementation \par+ Takes application specificities into account \par+ Reduces application workload & -- Specific implementation \par-- Manually implemented \\ \cline{2-5}
    & Mid-tier & -- Content processed between source of information and application & + Specific and inflexible implementation \par+ Automatically admits content \par+ Automatically maintains consistency \par+ Scales up complex components & -- Limited reusability \par-- Not fully transparent to developers \par-- A hit provides a small gain
    \\ \cline{2-5} 
    & Database & -- Partial or full tables \par-- Queries \par-- Views & + Improves performance and scalability of databases \par+ Cached data is provided to multiple applications \par+ Higher hit ratio \par+ Fully transparent to developers \par+ Easier to maintain consistency & -- Specific implementation \par-- A hit provides a small gain \\ \hline
\end{tabular}
\label{tab:cacheclassification}
\end{table}

Due to the variety of application domains, workload characteristics, and access patterns, there is no universal best caching solution, because it is hard, if not impossible, to achieve a deployment scheme that performs best in all environments and maintains its performance over an extended period. Typically, caching at lower levels (i.e.\ near the source of information, such as at hardware or database level) tends to achieve higher hit ratios, whilst caching at higher levels (i.e.\ near to where information is used or presented, such as client-side file caching) offers greater gains if a hit occurs, despite the expected lower hit ratios~\cite{Amza2005}. Therefore, caching at different locations is complimentary, and a combination of different solutions is possible and can bring even more benefits with the cost of a higher effort invested in caching configuration~\cite{Ravi2009,Sivasubramanian2007}.

Regardless of the cache location, coordination techniques should be employed to allow caching to improve the web-based system performance. Coordination topics are explored next.

\subsection{Coordination}
\label{subsec:coordination}

As discussed above, caches may be placed in several locations and, regardless of where and how the cache is implemented, issues such as the limited cache space should be addressed to ensure the cache efficiency. Cache coordination decisions address these issues. It includes deciding the adequate content to be cached, policies to avoid stale content, and how to deal with situations where the cache gets full of content, and there is a new content item to be added. Such strategies can be considered in any caching location. Coordination issues and associated strategies can be split into three main groups, namely admission, consistency, and replacement, which are detailed next.

\begin{description}

\item[Admission] Admission policies are used for content selection and preventing caching of unimportant content. They can have two key goals: (a) to prevent unnecessary items from being inserted into the cache~\cite{Baeza-Yate2007,Einziger2014}, and (b) to predict content expected to be requested and insert them into the cache in advance~\cite{Sulaiman2009}. The former is associated with the adoption of a \emph{reactive} mechanism to filter content and admitting only valuable items in the cache. The latter proactively prefetches and caches content expected to be requested, which can be seen as a \emph{proactive} admission policy, improving significantly cache performance and reducing the user perceived latency. Prefetching usually exploits the content spatial locality (e.g.\ correlated reference between content) and takes advantage of component's (server, proxy or client) idle time, preventing bandwidth underutilization and hiding part of the latency~\cite{Gawade2012,VeenaSinghBhadauriyaBhupeshGour2013}. \citeN{Ali2011} provided a comparison of server, client, and proxy prefetching challenges and issues. They also surveyed prefetching studies at the proxy level, and classified these studies into two broad categories (content-based and history-based), according to the data used for prediction.

\item[Consistency] Approaches that address consistency deal with the fact that once the content goes to the cache, it is potentially stale because cached items can be updated at their source. Therefore, if there are freshness requirements, it is necessary to design and implement consistency approaches to avoid staleness of cached items. There are two main consistency models to caching: (a) \emph{weak}, which provides higher availability by relaxing consistency (i.e.\ stale data is allowed to be returned from the cache)~\cite{Gupta2011}; and (b) \emph{strong}, which ensures that stale data is never returned, with a higher processing cost~\cite{Ports2010,Amza2005}. Both models can be achieved through validation or invalidation processes. With validation, the application explicitly invalidates cached content when they are modified by verifying the cached items in their source, e.g.\ with the origin server. With invalidation, cached content is evicted or updated as soon as changes in the underlying source of information are detected, or the source server notifies the cache system that a cached content has changed, in which strong consistency is not guaranteed. Furthermore, if weak consistency is adopted, timeouts---i.e.\ time to live (TTL)---are used to trigger evictions of expired cached content. Traditionally, consistency (or coherence) is treated as a separate issue from web caching, being widely studied, including in the area of distributed systems~\cite{Ghandeharizadeh2012a,Sivasubramanian2007,Gao2005}.

\item[Replacement] In situations in which the cache is full of content, replacement policies are responsible for deciding which one should be removed in case of new content should be added. Such policies ideally remove content that is less expected to result in hits in the future. To achieve such behavior, most of the proposed replacement policies associates a value (e.g.\ recency, frequency or worthiness) with each cache entry. When the cache achieves its full capacity and a new item should be inserted, entries are sorted according to their associated value. Then, the entry with the lowest value is removed. Traditional replacement strategies are classically classified into four categories: recency-based, frequency-based, function-based and randomized~\cite{Podlipnig2003,Ali2011}. Recently, the availability of monitoring workload and user accesses, and the dissemination of mechanisms to store and process such information have motivated the proposal of intelligent replacement policies, which exploit such data as training data to build a meta-model~\cite{Ali2011}. Despite that the most popular policy is the \emph{least recently used}, due to its simplicity and relatively good performance, several other replacement policies have been proposed with the aim of getting good performance in many situations~\cite{Wong2006}, and almost all of them were demonstrated to be superior to others in their proposal~\cite{Podlipnig2003,AliAhmed2011,Sulaiman2013}. These results indicate that there is no best policy in all scenarios because a policy performance depends on workload characteristics. Focusing on these differences, \citeN{Wong2006} provides a pragmatic approach to choosing the best policy. In summary, each policy imposes a trade-off between success rate and computational overhead.

\end{description}

Many factors can be observed while choosing or proposing a coordination strategy. Coordination strategies usually evaluate temporal locality properties, such as recency (i.e.\ last time a piece of content was requested) and frequency (i.e.\ how many requests to a piece of content occurred)~\cite{AliAhmed2011}. In addition to temporal locality, the spatial locality can also be explored by analyzing the content context, which is not always straightforward. Other factors include size, access latency and time since last modification of the content, or even heuristics, such as useful time of a cached item. However, taking into account all factors to achieve an improved coordination decision is not a trivial task, given the computational overhead. Furthermore, there are no well-accepted influence factors, because all of them are subject to the particularities of the environment requirements, which means that the importance of properties depends on the scenario or environment under consideration~\cite{Podlipnig2003,Wong2006}.

In summary, all coordination strategies aim to solve caching issues to improve the general performance of web-based applications. However, cache benefits come with the cost of investing an effort to design, implement, and maintain the cache. The benefits of each caching strategy can be estimated regarding improvements to the performance of the cache and the system, which is measured with different metrics that are described next.

\subsection{Measurement}
\label{subsec:measurement}

The last category of our taxonomy for web caching is related to measurement techniques, which are used to measure the efficiency of web caching strategies~\cite{Podlipnig2003,Ali2011}. The most popular and well-accepted metrics are summarized in Table~\ref{tab:metrics} with their name and acronym, description and definition of the metric. Similarly to coordination strategies presented in the previous section, cache performance metrics can be applied in any caching scheme, being not limited to web caching. Furthermore, these metrics are well-known and generic enough to evaluate and compare any caching strategy and tuning task.

\begin{table}
\scriptsize
\centering
\caption{Traditional Cache Evaluation Metrics.}%
\begin{tabular}{|p{2.5cm}|p{8.7cm}|c|}
\hline
\textbf{Metric (Acronym)} & \textbf{Description} & \textbf{Definition} \\ \hline
\textbf{Hit Ratio (HR)} & The fraction of requests that result in a cache hit. Usually, the higher the hit ratio, the better the technique employed. & \multirow{2}{*}{$\sum _{i \epsilon R} \frac{h_{i}}{f_{i}}$} \\ \hline
\textbf{Byte Hit Ratio (BHR)} & The fraction of bytes of requests that result in a cache hit over the total number of bytes requested by clients. & \multirow{2}{*}{$\sum _{i \epsilon R} s_{i} \cdot \frac{h_{i}}{f_{i}}$} \\ \hline
\textbf{Latency Savings Ratio (LSR)} & The approximate response time reduction provided by caching while processing client requests. & \multirow{2}{*}{$\sum _{i \epsilon R} d_{i} \cdot \frac{h_{i}}{f_{i}}$} \\ \hline
\textbf{Throughput (TRP)} & The number of requests processed per second throughout a period of time. A higher throughput shows the effectiveness of the caching, as more requests can be satisfied within the same period of time. & \multirow{2}{*}{$\sum _{i \epsilon R} \frac{f_{i}}{t_{i}}$} \\ \hline
\multicolumn{3}{|l|}{\textbf{Notation:}} \\
\multicolumn{3}{|l|}{$s_{i} =$ size of item i} \\
\multicolumn{3}{|l|}{$f_{i} =$ total number of requests for item i} \\
\multicolumn{3}{|l|}{$h_{i} =$ total number of hits for item i} \\
\multicolumn{3}{|l|}{$t_{i} =$ total fetch delay in seconds of requests for item i} \\
\multicolumn{3}{|l|}{$d_{i} =$ mean fetch delay from server for item i} \\
\multicolumn{3}{|l|}{$R =$ set of all requests} \\ \hline
\end{tabular}
\label{tab:metrics}
\end{table}

Note that HR and BHR conflict with each other, given that keeping small popular items in the cache favors HR, whereas holding large popular items increases BHR. Moreover, because it is hard to measure the response time precisely (it is affected by many factors independent from the cache, such as network congestion and server stability), LSR and TRP require a well-specified performance test~\cite{Wong2006}.

Proactive admission strategies, which focus on predicting web items expected to be requested in the future, require specific metrics to evaluate efficiency and performance of these strategies. \citeN{Domenech2006} reviewed a representative subset of the prediction algorithms used specifically for web prefetching, proposing a taxonomy based on three categories (prediction-related, resource usage and latency-related), which allow us to identify analogies and differences among the commonly used measurements. Table~\ref{tab:predictionmetrics} presents such prediction-based metrics~\cite{Domenech2006,Huang2008}.

\begin{table}
\scriptsize
\centering
\caption{Prediction-based Cache Evaluation Metrics.}%
\begin{tabular}{|l|p{7.5cm}|c|}
\hline
\textbf{Metric} & \textbf{Description} & \textbf{Definition} \\ \hline
\textbf{Precision} & The fraction of prefetched requests that subsequently result in a cache hit. & \multirow{2}{*}{$\frac{GoodPredictions}{Predictions}$} \\[2ex] \hline
\textbf{Byte Precision} & The fraction of prefetched bytes of requests that subsequently result in a cache hit. & \multirow{2}{*}{$\frac{SizeOfGoodPredictions}{Predictions}$} \\[2ex] \hline
\textbf{Recall} & The fraction of prefetched requests that subsequently result in a cache hit over the total number of requests. & \multirow{2}{*}{$\frac{GoodPredictions}{UserRequests}$} \\[2ex] \hline
\textbf{Byte Recall} & The fraction of prefetched bytes of requests that subsequently result in a cache hit over the total number of bytes requested. & \multirow{2}{*}{$\frac{SizeOfGoodPredictions}{UserRequests}$} \\ \hline
\multicolumn{3}{|l|}{\textbf{Notation:}} \\
\multicolumn{3}{|l|}{$GoodPredictions =$ the amount of predicted items that are subsequently demanded by the user} \\
\multicolumn{3}{|l|}{$Predictions =$ the amount of items predicted by the prediction algorithm} \\
\multicolumn{3}{|l|}{$SizeOfGoodPredictions =$ the number of $GoodPredictions$ with their size in bytes} \\
\multicolumn{3}{|l|}{$UserRequests =$ the amount of items that the user demands} \\ \hline
\end{tabular}
\label{tab:predictionmetrics}
\end{table}
\end{appendices}

\end{document}